\documentclass[twocolumn,showpacs,epsf,preprintnumbers,amsmath,amssymb]{revtex4}

\usepackage{graphicx}
\usepackage{dcolumn}
\usepackage{bm}

\begin{document}
\title{Entanglement Preserving in Quantum Copying of Three-qubit Entangled State}
\author{TONG Zhao-Yang and KUANG Le-Man }
\address{ Department of Physics, Hunan Normal University, Changsha 410081, China}

\begin{abstract}
We study the degree to which quantum entanglement survives when a
three-qubit entangled state is copied by using  local and
non-local processes, respectively, and investigate iterating
quantum copying for the three-qubit system. There may exist
inter-three-qubit entanglement and inter-two-qubit entanglement
for the three-qubit system.  We show that both local and non-local
copying processes degrade quantum entanglement in the
three-particle system due to a residual correlation between the
copied output and the copying machine. We also show that the
inter-two-qubit entanglement is preserved better than the
inter-three-qubit entanglement in the local cloning process. We
find that non-local cloning is much more efficient than the local
copying for broadcasting entanglement, and output state via
non-local cloning exhibits the fidelity better than local cloning.
\end{abstract}


\maketitle

\section{Introduction}      

The most fundamental difference between classical and quantum
information is that while classical information can be copied
perfectly, quantum information cannot. In particular, it follows
from the no-cloning theorem \cite{1}  that one cannot create a
perfect duplicate of an arbitrary qubit. Nevertheless, Buzek and
Hillery and other authors \cite{2,3,4,5} have shown that imperfect
copies can be made by a universal quantum cloning machine (UQCM),
the outputs of which are identical. The price which must be paid
is that there is a difference between the original input and the
copies, because of residual entanglement between the machine and
copies. However, not only similarity is lost during the cloning
process. Perhaps even more import than the no-cloning feature of
quantum mechanics is entanglement, first noted by
Einstein-Podolsky-Rosen (EPR)and  Schr\"{o}dinger \cite{6}. For
decades, quantum entanglement has been the focus of much work in
the foundations of quantum mechanics. In particular, it is
associated with quantum nonseparability, the violation of Bell's
inequalities, and the so-called EPR paradox. Beyond this
fundamental aspect, creating and manipulating of entangled states
are essential for quantum information applications. Among these
applications are quantum teleportation \cite{7}, quantum dense
coding \cite{8}, quantum error correction \cite{9}, and quantum
computational speedups \cite{10}, quantum cryptography \cite{10'},
and quantum positioning and clock synchronization \cite{10''}.
Hence, quantum entanglement has been viewed as an essential
resource for quantum information processing.  all of these
applications depend upon the strength of quantum entanglement. If
the cloning process applied to entangled subsystems is to be
anything more than a basic curiosity, and is to find a practical
application in the field of quantum information theory, it should
be possible to obtain not only maximally accurate copies of the
original state, but also copies which preserve entangling
characteristic of the original state when the input state is an
entangled state. Masiak and Knight \cite{11} have shown that
copies of entangled pair of qubits can be generated by using the
UQCM. In this paper, We will investigate the degree to which
entanglement survives when a three qubit entangled state is copied
by using  local and non-local processes, respectively. We will
show how these copying processes degrade quantum entanglement due
to a residual correlation between the copied output and the
copying machine. We will also show that entanglement is rapidly
destroyed by the copying process.

\section{Entanglement variation and state fidelity in quantum copying}
We consider quantum copying of a non-local state of a
three-particle system consisting of a three qubits and assume that
the three qubits are initially  prepared in a entangled pure state
expressed by
\begin{equation}
\label{1}
\mid\psi\rangle =\cos\alpha\mid 000\rangle
+\sin\alpha\mid111\rangle.
\end{equation}

There are two approaches to realize quantum copying of the above
entangled state. The first one is that each of the three original
qubits is copied separately by using three different local copying
machines. The second method is that the entangled state of the
three qubits is treated as a state in a large Hilbert space and
copied as a whole. The former is called as local cloning while the
latter is regarded as non-local cloning.

We will measure quantum entanglement in a three-particle system by
making use of the entanglement tensor approach which is first
proposed in Ref. \cite{13'} and further developed for a
three-particle system by present authors \cite{13}. Entanglement
measures of a three-qubit system involve both an inter-three-qubit
entanglement measure denoted by $E_3$ and an inter-two-qubit
entanglement measure labelled by $E_2$. Based on entanglement
tensors  $M(1,2,3)$ and $M(m,n)(1\leq m<n\leq 3)$ introduced in
Ref. \cite{13'}, entanglement measures  $E_3$ and $E_2$ can be
defined as follows
\begin{eqnarray}
\label{2}
E_3=\frac{1}{4}\sum_{i,j,k=1}^3M_{ijk}(1,2,3)M_{ijk}(1,2,3), \\
\label{3}
E_2(m,n)=\frac{1}{3}\sum_{i,j=1}^3M_{ij}(m,n)M_{ij}(m,n),
\end{eqnarray}
Here the entanglement tensors on the RHS of Eqs. (\ref{2}) and
(\ref{3}) are defined by
\begin{eqnarray}
\label{4}
M_{ij}(m,n)&=&K_{ij}(m,n)-\lambda_{i}(m)\lambda_j(n),\\
\label{5}
M_{ijk}(1,2,3)&=&K_{ijk}(1,2,3)-\lambda_i(1)M_{jk}(2,3)\nonumber \\
&&-\lambda_j(2)M_{ik}(1,3)-\lambda_k(3)M_{ij}(1,2)\nonumber \\
&&-\lambda_i(1)\lambda_j(2)\lambda_k(3),
\end{eqnarray}
where $K_{ijk}(1,2,3)$, $K_{ij}(m,n)$, and $\lambda_{i}(m)$ are
the correlation tensors and coherence vectors defined in
Ref.\cite{13}, respectively.

These measures are invariant under local unitary transformations
of the subsystems and their values can change between $0$
(unentangled states) and $1$ (maximum entangled states). $E_3$
quantifies the three-qubit entanglement . The larger $E_3$ is, the
stronger the three-qubit entanglement is. And $E_2(m,n)$
quantifies the entanglement between any two qubits $m,n$ in the
three-qubit system. The larger $E_2(m,n)$ is, the stronger  the
entanglement between two qubit $m,n$ is. Making use of Eqs. (2-8)
in Ref.\cite{13} we can obtain the nonzero components of the
 coherence vector and correlation tensor
\begin{eqnarray}
\label{e6}
\lambda_3(1)&=&\lambda_3(2)=\lambda_3(3)=-\cos(2\alpha),
\\
\label{e7}
K_{33}(1,2)&=&K_{33}(2,3)=K_{33}(1,3)=1, \\
\label{e8} K_{111}&=&\sin(2\alpha),
          \hspace{0.3cm}K_{333}=-\cos(2\alpha),\\
\label{e9} K_{122}&=&K_{212}=K_{221}=-\sin(2\alpha).
\end{eqnarray}

From Eqs. (\ref{4}) and (\ref{5}) one can find  the nonzero
components of the entanglement tensor
\begin{eqnarray}
\label{e10}
M_{111}&=&\sin(2\alpha),\hspace{0.3cm}\\
\label{e11}
M_{333}&=&2\sin^2(2\alpha)\cos(2\alpha), \\
\label{e12}
M_{122}&=&M_{212}=M_{221}=-\sin(2\alpha),\\
\label{e13} M_{33}(1,2)&=&M_{33}(2,3)=M_{33}(1,3)\nonumber \\
&=&\sin^2(2\alpha),
\end{eqnarray}
while all other components of the entanglement tensor vanish. Then
from Eqs. (\ref{2}) and (\ref{3})we find the values of
entanglement measures
 $E_{3}$ and $E_{2}(m,n)$ as follows
\begin{eqnarray}
\label{e14}
E_{3}&=&\sin^{2}(2\alpha)\left (1+\sin^{2}2\alpha \cos^{2}(2\alpha)\right ),  \\
\label{e15}
E_{2}(1,2)&=&E_{2}(2,3)=E_{2}(1,3)=\frac{1}{3}\sin^{4}(2\alpha),
\end{eqnarray}
which indicate that the three-qubit GHZ state $\mid\psi\rangle
=(\mid 000\rangle +\mid111\rangle)/\sqrt{2}$ is the maximally
entangled three-qubit state with the maximal value of the
inter-three-qubit entanglement $E_3=1$.

It is well known that there are two approaches to realize quantum
copying of the entangled state given by Eq. (\ref{1}). The first
one is that each of the three original qubits is copied separately
by using three different local copying machines. The second method
is that the entangled state of the three qubits is treated as a
state in a large Hilbert space and copied as a whole. The former
is called as local cloning while the latter is regarded as
non-local cloning. In what follows we will investigate
entanglement variation and copying fidelity in the two types of
quantum cloning processes.

\begin{center}
{\bf A. Local cloning}
\end{center}
A scheme which will achieve local cloning is described by the
following process \cite{12}: three distant parties share an
entangled three qubit state $\mid \psi\rangle$. Each of them
performs some local transformations on their own qubit using
distant quantum copying machines. We assume that the three
additional qubits, employed in the copying process, are initially
unentangled. Three copiers separately make copies of the qubits by
the following local unitary transformations \cite{17}
\begin{eqnarray}
U_1\mid 0\rangle_{a_0}\mid 0\rangle_{a_1}\mid
X\rangle_x&=&\sqrt{\frac{2}{3}}
\mid 00\rangle_{a_0a_1}\mid \uparrow\rangle_x  \nonumber \\
& &+\sqrt{\frac{1}{3}}\mid +\rangle_{a_0a_1}\mid \downarrow\rangle_x, \\
U_1\mid 1\rangle_{a_0}\mid 0\rangle_{a_1}\mid
X\rangle_x&=&\sqrt{\frac{2}{3}}
\mid 11\rangle_{a_0a_1}\mid \downarrow\rangle_x \nonumber \\
& &+\sqrt{\frac{1}{3}}\mid +\rangle_{a_0a_1}\mid \uparrow\rangle_x
\end{eqnarray}
where
$\mid\!+\!\rangle_{a_0a_1}\!=\!(\mid\!10\rangle_{a_0a_1}\!+\!\mid\!01\rangle_{a_0a_1})
/\!\sqrt{2}$. The system labelled by $a_0$ is the original (input)
qubit, while the other system $a_1$ represents the target qubit
onto which the information is copied. The states of the copying
machine are labelled by $x$. The state space of the copying
machine is two dimensional and we assume that it is always in the
same state $\mid X\rangle_x$ initially. The result of the cloning
process is an output state which is no longer a pure state, but is
a mixed state described by the following density matrix
\begin{eqnarray}
\label{6}
\hat{\rho}&=&\frac{1+124\cos^2\alpha}{216}\mid
000\rangle\langle
000\mid\nonumber \\
&&+\frac{1+124\sin^2\alpha}{216}
\mid 111\rangle\langle 111\mid\nonumber \\
& &+\frac{8\sin\alpha \cos\alpha}{27}(\mid 111\rangle\langle
000\mid
+\mid 000\rangle\langle 111\mid)\nonumber \\
& &+\frac{5+20\sin^2\alpha}{216}(\mid 110\rangle\langle 110\mid
+\mid 011\rangle\langle 011\mid \nonumber \\
& &+\mid 101\rangle \langle 101\mid )+\frac{5+20\cos^2\alpha}{216}
 (\mid 100\rangle\langle 100\mid \nonumber \\
& &+\mid 010\rangle\langle 010\mid +\mid 001\rangle\langle
001\mid).
\end{eqnarray}

We now calculate entanglement for the output state of local
quantum copying described by above density operator. Making use of
Eqs. (2-8) in Ref.\cite{13} we can obtain the nonzero coherence
vectors and correlation tensors
\begin{eqnarray}
\label{7}
\lambda_3(1)&=&\lambda_3(2)=\lambda_3(3)=-\frac{2}{3}\cos(2\alpha),
\\
\label{8}
K_{33}(1,2)&=&K_{33}(2,3)=K_{33}(1,3)=\frac{4}{9}, \\
\label{9}
K_{111}&=&\frac{8}{27}\sin(2\alpha),
          \hspace{0.3cm}K_{333}=\frac{8}{27}\cos(2\alpha),\\
\label{10} K_{122}&=&K_{212}=K_{221}=-\frac{8}{27}\sin(2\alpha).
\end{eqnarray}
From Eqs. (\ref{4}) and (\ref{5}) one can find  the following
entanglement tensors
\begin{eqnarray}
\label{11}
M_{111}&=&\frac{8}{27}\sin(2\alpha),\hspace{0.3cm}\\
\label{12}
M_{333}&=&\frac{16}{27}\sin^2(2\alpha)\cos(2\alpha), \\
\label{13}
M_{122}&=&M_{212}=M_{221}=-\frac{8}{27}\sin(2\alpha),\\
\label{14} M_{33}(1,2)&=&M_{33}(2,3)=M_{33}(1,3)\nonumber \\
&=&\frac{4}{9}\sin^2(2\alpha),
\end{eqnarray}
while all other components of the entanglement tensor  vanish.
Then from Eqs. (\ref{2}) and (\ref{3})we find the values of
$E_{3}$ and $E_{2}(m,n)$ entanglements as follows
\begin{eqnarray}
\label{15}
E_{3}&=&\frac{64}{729}\sin^{2}(2\alpha)\left (1+\sin^{2}2\alpha \cos^{2}(2\alpha)\right ),  \\
\label{16}
E_{2}(1,2)&=&E_{2}(2,3)=E_{2}(1,3)=\frac{16}{243}\sin^{4}(2\alpha),
\end{eqnarray}
which indicate that both the inter-three-qubit entanglement and
inter-two-qubit entanglement in a three qubit system characterized
by $E_3$ and $E_{2}(m,n)$, respectively, can be broadcasted via
local quantum cloning, since the amount of  $E_{3}$ and
$E_{2}(m,n)$ entanglements are nonzero, i.e.,  $E_{3}\neq 0$,
$E_{2}(m,n)\neq 0$ except $\alpha = 0$ or $\alpha=\frac{\pi}{2}$.
In particular, when $\alpha=\pi/4$, we have $E_3=64/729$ and
$E_2(n,m)=16/243$ which recover the result in Ref. \cite{13} for
quantum copying of the GHZ state.

Comparing remaining $E_3$ and $E_{2}(m,n)$ entanglements after
local cloning  given by Eqs. (\ref{25}) and (\ref{26}) with the
original $E_3$ and $E_{2}(m,n)$ entanglements before local cloning
given by Eqs. (\ref{e14}) and (\ref{e15}), we can find that both
$E_3$ and $E_{2}(m,n)$ entanglements are reduced in the local
cloning process. The remaining $E_3$ entanglement after local
cloning is only 8.76 percent of the original $E_3$ entanglement,
and the remaining $E_2$ entanglement after local cloning is  19.8
percent of the original $E_2$ entanglement. Therefore, the  $E_2$
entanglement is preserved better than the  $E_3$ entanglement in
the local cloning process.

\begin{figure}
\begin{center}
\includegraphics[width=3.8in,height=2.6in]{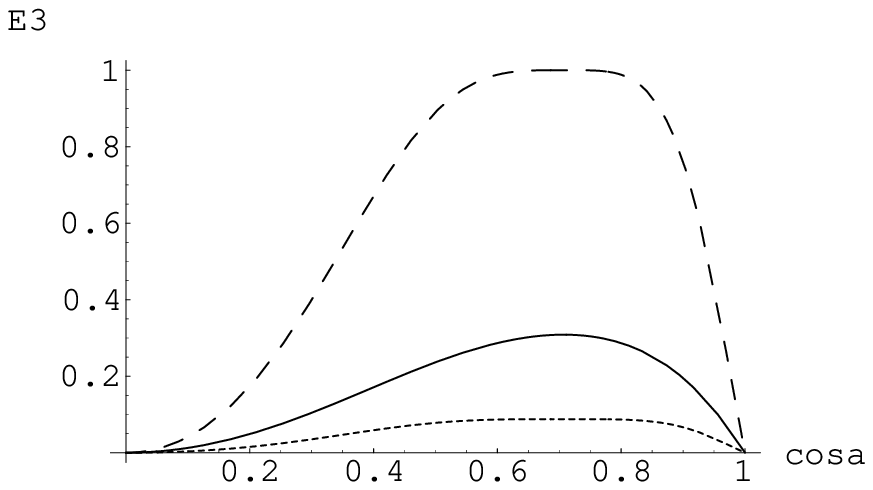}
\end{center}
\caption{$E_3$ entanglement  of the three-qubit pure state
$\mid\psi\rangle$ (dashed curve) and  remaining $E_3$ entanglement
after the first step of local (dotted curve) and non-local (solid
curve) cloning.}
\end{figure}

\begin{center}
{\bf B. Non-local cloning}
\end{center}
The entangled state given in Eq. (\ref{3})  $\mid\psi\rangle$ can
also be cloned non-locally \cite{14}. In this case the entangled
state of the three-qubits is treated as a state in a larger
Hilbert space and cloned as a whole. The non-local quantum copying
machine  \cite{18} is an $N$ dimensional quantum system, and we
shall let $\mid X_i\rangle_x$ ($i=1,\cdots,N$) be a set of
orthonormal basis of the copying machine Hilbert space. This
copier is initially prepared in a particular state $\mid
X\rangle_x$. The action of the cloning transformation can be
specified by a unitary transformation acting on the basis vectors
of the tensor product space of the original quantum system $\mid
\phi_i\rangle_{a_0}$, the copier, and an additional
$N$-dimensional system which is to become the copy (which is
initially prepared in an arbitrary state $\mid 0\rangle_{a_1}$).
The corresponding transformation $U_2$ is given by
\begin{eqnarray}
  U_2\mid\phi_i\rangle_{a_0}\mid 0\rangle_{a_1} \mid X\rangle_x
 &=&c \mid \phi_i \rangle_{a_0} \mid \phi_i \rangle_{a_1} \mid X_i \rangle_x  \nonumber \\
 & &+d\sum_{j\neq i}^N(\mid\phi_i \rangle_{a_0} \mid \phi_j \rangle_{a_1} \nonumber \\
 & &+ \mid \phi_j \rangle_{a_0} \mid \phi_i \rangle_{a_1}) \mid\!X_j\rangle_x,
\end{eqnarray}
where $ i=1,\cdots,N$, $c^2\!=\!2/(N+1)$,  and $d^2\!=\!1/2(N+1)$.

The final state of the three-qubit copies at the output of the
cloning machine is given by the following density operator
\begin{eqnarray}
\label{17}
\hat{\rho}&=&\frac{1+10\cos^2\alpha}{18}\mid 000\rangle\langle 000\mid\nonumber \\
& &+\frac{1+10\sin^2\alpha}{18}\mid 111\rangle\langle 111\mid\nonumber \\
& &+\frac{5\sin\alpha\cos\alpha}{9}(\mid 111\rangle\langle 000\mid
+\mid 000\rangle\langle 111\mid) \nonumber \\
& &+\frac{1}{18}(\mid 110\rangle\langle 110\mid +\mid
011\rangle\langle 011\mid +\mid 101\rangle \langle 101\mid \nonumber \\
& &+\mid 100\rangle\langle 100\mid +\mid 010\rangle\langle 010\mid
+\mid 001\rangle\langle 001\mid).
\end{eqnarray}
Now we check whether entanglement is broadcasted. Through lengthy
calculations, we obtain the nonzero components of the coherent
vector and correlation tensor
\begin{eqnarray}
\label{18}
\lambda_3(1)&=&\lambda_3(2)=\lambda_3(3)=-\frac{5}{9}\cos(2\alpha),
\\
\label{19}
K_{33}(1,2)&=&K_{33}(2,3)=K_{33}(1,3)=\frac{5}{9}, \\
\label{20}
K_{111}&=&\frac{5}{9}\sin(2\alpha),
          \hspace{0.3cm}K_{333}=-\frac{5}{9}\cos(2\alpha),\\
\label{21}
K_{122}&=&K_{212}=K_{221}=-\frac{5}{9}\sin(2\alpha),
\end{eqnarray}
which leads to the following non-vanishing components of
entanglement tensor
\begin{eqnarray}
\label{22}
M_{111}&=&\frac{5}{9}\sin(2\alpha),\\
\label{23}
M_{333}&=&\frac{10}{27}\left (1-\frac{25}{27}\cos^2(2\alpha)\right )\cos(2\alpha), \\
\label{24}
M_{122}&=&M_{212}=M_{221}=-\frac{5}{9}\sin(2\alpha),\\
\label{25}
M_{33}(1,2)&=&M_{33}(2,3)=M_{33}(1,3)\nonumber \\
&=&\frac{5}{9}-\frac{25}{81}\cos^2(2\alpha).
\end{eqnarray}

Substituting Eqs.(\ref{22}-\ref{25}) into Eqs. (\ref{2}) and
(\ref{3}), we find the values $E_{3}$ and $E_{2}(m,n)$ to be
\begin{eqnarray}
\label{26}
E_3&=&\frac{25}{81}\sin^{2}(2\alpha) \nonumber \\
&&+\frac{25}{729}\left (1-\frac{25}{27}\cos^{2}(2\alpha)\right
)^{2}\cos^{2}2\alpha,
\\
\label{27}
E_{2}(1,2)&=&E_{2}(2,3)=E_{2}(1,3)\nonumber\\
&=&\frac{25}{243}\left (1-\frac{5}{9}\cos^{2}(2\alpha)\right )^2.
\end{eqnarray}
From which we see that both $E_3$ and $E_2$ entanglements can be
broadcasted via the non-local process. Obviously, when
$\alpha=\pi/4$, we have $E_3=25/81$ and $E_2(n,m)=25/243$ which
recover the result in Ref. \cite{13} for quantum copying of the
GHZ state.

\begin{figure}
\begin{center}
\includegraphics[width=3.8in,height=2.6in]{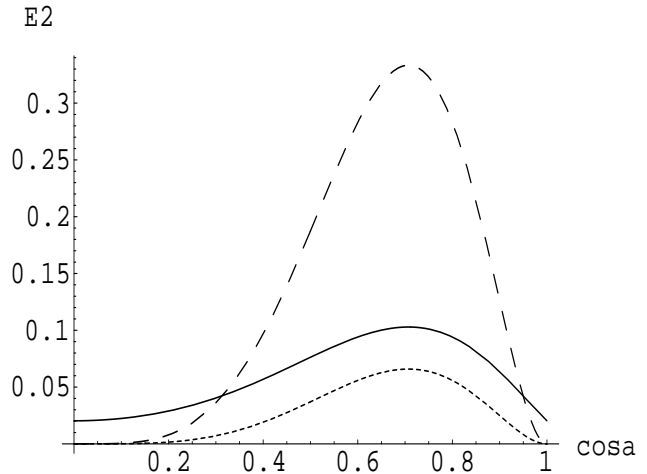}
\end{center}
\caption{$E_2$ entanglement of the three-qubit pure state
$\mid\psi\rangle$ (dashed curve) and  remaining $E_2$ entanglement
after the first step of local (dotted curve) and non-local (solid
curve) cloning.}
\end{figure}

We now numerically discuss  entanglement variation in the local
and non-local copying processes with respect to the values of
$\cos\alpha$. In Fig.1 and Fig.2 we plot the $E_3$ entanglement
and $E_2$ entanglement in the three-qubit input state
$\mid\psi\rangle$ and the three-qubit output states obtained as
the result of  local and non-local copying of the state
$\mid\psi\rangle$, respectively. From Fig.1 and Fig.2 we can see
that both local and non-local copy can broadcast entanglement in
the three-qubit system. And we find that non-local cloning is much
more efficient than the local copying process for broadcasting
entanglement. In the local cloning case, the amount of both $E_3$
entanglement and $E_2$ entanglement of the copied state is less
than that of the original state for all values of $\cos\alpha$. In
the non-local cloning case, the amount of $E_3$ entanglement  of
the copied state is always less than that of the original state
for all values of $\cos\alpha$ while the amount of $E_2$
entanglement of the copied state is  less than that of the
original state for most range of $\cos\alpha$, i.e.,
$0.33065<\cos\alpha<0.95287$. In other words, the remaining $E_2$
entanglement after the non-local cloning is amplified versus the
initial state when $\cos\alpha<0.33065$ or $\cos\alpha>0.95287$.
This inter-two-particle entanglement amplification in a
three-particle system is originated from inter-two-particle, and
particle-copier interaction. However, inter-three-qubit
entanglement can not be amplified for both local and nonlocal
cloning processes. Hence this does not violate the well known fact
that copying a quantum system does not increase the information
obtainable about the originals.

Finally, It is interesting to compare the fidelity $F_1$ of the
output density operator after local copying relative to
$\mid\psi\rangle$ with the fidelity $F_2$ of the output density
operator after non-local copying relative to $\mid\psi\rangle$.
The fidelity of a density matrix $\rho$ relative to
$\mid\psi\rangle$ is defined by
$F=\langle\psi\mid\rho\mid\psi\rangle$. From Eqs. (\ref{1}),
(\ref{e6}), and (\ref{17}) we get that
\begin{eqnarray}
\label{28}
F_1&=&\frac{125}{216}-\frac{15}{27}\sin^{2}\alpha \cos^{2}\alpha,\\
\label{29}
F_2&=&\frac{11}{18}.
\end{eqnarray}
From which we find that the fidelity $F_2$ of the output state
after non-local copying is independent of the values of
$\cos\alpha$. This means that whether the initial state is like,
the fidelity of the output is definite. However,  in the local
copying process, the fidelity $F_1$ depends upon the initial
state. From Eq. (\ref{28}), we see that when $\cos\alpha =0$ or
$1$, the output state is the closest to the original state. When
$\cos\alpha=\sqrt{2}/2$, the output state is the farthest to the
original state. In Fig.3 we plot the fidelity with respect to
$\cos\alpha$ for local and nonlocal cloning cases, respectively.
From Fig. 3 it can be seen that the fidelity in the non-local
cloning process is always greater than  that in the local cloning
process, i.e., $F_2>F_1$. This means that the output state after
non-local copying is closer to the original state than the output
state after local copying.

\begin{figure}
\begin{center}
\includegraphics[width=3.8in,height=2.6in]{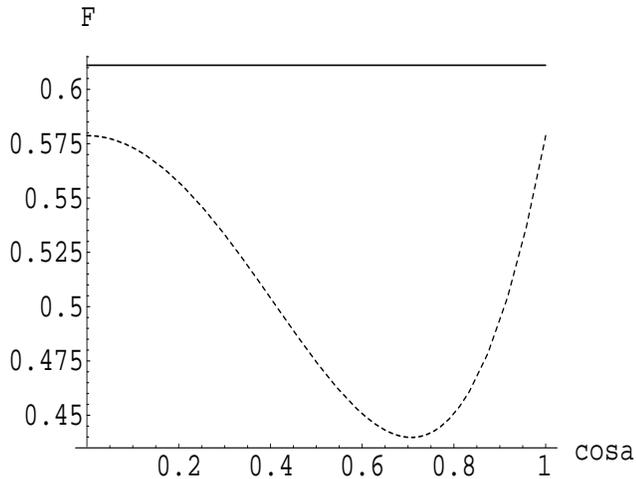}
\end{center}
\caption{Fidelity of the output state after local cloning (dotted
curve) and  non-local cloning (solid curve).}
\end{figure}

\section{Repetition of Non-local Cloning}
We can see from the previous section that both local and non-local
copying are not rather efficient process from the point of view of
preserving entanglement. Even in the case when the maximally
inseparable state, i.e. GHZ state where $\alpha=\pi/4$, is the
input state of the cloning process, only a small amount of
entanglement survives the cloning. It is an interesting question
to ask what will happen when the output state of the copier is
used as an input state in the next step of a sequence of cloning
process. In particular, we are interested in  discovering how fast
the entanglement decreases when the copying is iterated. We
restrict ourself in the case of non-local cloning, because this
scheme is much more effective as we saw above, and results
obtained in this case can be treated as an upper bound for all
other schemes.

The final state of the copying process is a mixed state, described
by the density matrix, Eq. (\ref{17}). This density matrix cannot
be used directly as input data in computations, because the
non-local copying scheme works straightforwardly only when an
input state is initially in a pure, potentially entangled state.
The density matrix should be first converted to a form which
allows us to perform the second cloning.  It turns out that a
simple diagonalization of the density matrix is sufficient. In
this new diagonal basis, the density matrix is given by the
mixture of projection operators
\begin{equation}
\label{30}
\hat{\rho}=\sum_{j=1}^{8}\alpha_{j}\mid\varphi_{j}\rangle\langle\varphi_{j}\mid.
\end{equation}
The weights $\alpha_{j}$ in the decomposition are the eigenvalues
of the density matrix $\hat{\rho}$ and the vectors
$\mid\varphi_{j}\rangle$ are the normalized eigenvectors of
$\hat{\rho}$. Each vector $\mid\varphi_{j}\rangle$ can be cloned
separately. The mixture of the resultant density matrices taken
with the weights $\alpha_{j}$ is the result of the second cloning
process.

As a specific example of iterating copying, in what follows we
consider repetition of non-local quantum cloning for the GHZ state
given by Eq. (\ref{1}) through taking $\alpha=\pi/4$. After the
first non-local cloning, from Eq. (\ref{17}) we express the
density operator of the out state of the original three input
qubits as the form of the state (\ref{30} with the coefficients
\begin{equation}
\label{31}
 \alpha_8=\frac{11}{18}, \hspace{0.3cm}
 \alpha_i=\frac{1}{18} \hspace{0.2cm}(i=1,2, \cdots, 7),
\end{equation}
and eigenvectors of the out-state density operator given by
\begin{eqnarray}
\label{32}
\mid\varphi_{1}\rangle&=&\frac{1}{\sqrt2}(\mid000\rangle-\mid111\rangle),\hspace{0.3cm}
                     \mid\varphi_{2}\rangle=\mid110\rangle,\\
\mid\varphi_{3}\rangle&=&\mid101\rangle, \hspace{0.2cm}
\mid\varphi_{4}\rangle=\mid100\rangle,\\
\mid\varphi_{5}\rangle&=&\mid011\rangle,
\hspace{0.2cm}\mid\varphi_{6}\rangle=\mid010\rangle,\\
\mid\varphi_{8}\rangle&=&\frac{1}{\sqrt2}(\mid000\rangle+\mid111\rangle),
\hspace{0.3cm}\mid\varphi_{7}\rangle=\mid001\rangle.
\end{eqnarray}

\begin{table}[h]
\begin{center}
\begin{tabular}{|c|c|c|c|c|c|c|c|}
\hline
{\sl No} & 0 & 1 & 2&3&4&5&6 \\
\hline
$E_{3}$&1.0000&0.3086&0.0953&0.0294&0.0091&0.0028&0.0000 \\
\hline
$E_{2}$&0.3333&0.1029&0.0318&0.0098&0.0030&0.0009&0.0000\\
\hline
\end{tabular}
\end{center}
\caption{Entanglement $E_{3}$ and $E_{2}$ of clones of the GHZ
state as a function of the number of cloning steps.}
\end{table}
The state described by the density operator (\ref{30}) is the
input state of the second copying. We now separately copy the
eigenstates of the density operator (\ref{30}), i.e.,
$\varphi_{i}\rangle$, via non-local process. After the second
copying, we arrive at the output state expressed by the following
density operator
\begin{eqnarray}
\label{33}
\hat{\rho}&=&\frac{54}{13}(\mid 000\rangle\langle 000\mid+\mid 111\rangle\langle 111\mid) \nonumber \\
& &+\frac{25}{162}(\mid 111\rangle\langle 000\mid
+\mid 000\rangle\langle 111\mid) \nonumber \\
& &+\frac{7}{81}(\mid 110\rangle\langle 110\mid +\mid
011\rangle\langle 011\mid +\mid 101\rangle \langle 101\mid \nonumber \\
& &+\mid 100\rangle\langle 100\mid +\mid 010\rangle\langle 010\mid
+\mid 001\rangle\langle 001\mid),
\end{eqnarray}
which leads to the following entanglement tensors
\begin{eqnarray}
\label{34}
M_{122}=&-M_{111}=M_{212}=M_{221}=-\frac{25}{81},\\
M_{33}(1,2)=&M_{33}(2,3)=M_{33}(1,3)=\frac{25}{81}.
\end{eqnarray}
Hence after finishing the second copying, entanglement of the
three-qubit system  becomes
\begin{eqnarray}
\label{35}
E_3&=&\left (\frac{25}{81}\right )^2, \nonumber \\
\label{27} E_{2}(1,2)&=&E_{2}(2,3)=E_{2}(1,3)=\frac{1}{3}\left
(\frac{25}{81}\right )^2.
\end{eqnarray}

From above calculations we can see that entanglement in the
three-qubit system is  further reduced after the second copying.
Repeating the above procedure we can investigate changes in
entanglement $E_3$ and $E_2$ at each stage of iterating copy
process. In Table 1 we show the entanglement measures $E_{3}$ and
$E_{2}(m,n)$ of the clones of the maximally entangled GHZ state
for the case of the first five stages of the cloning. From the
this table, we see that the entanglement decreases extremely
rapidly, and after just six iterations the copy of the GHZ state,
the resultant entanglement goes to zero.

\section{Conclusions}

In conclusion,  we have investigated remaining entanglement  in
quantum copying and iterating quantum copying of a three-qubit
system by using the entanglement tensor approach. Specifically, we
have studied the remaining inter-two-qubit and inter-three-qubit
entanglements in the local and non-local cloning processes,
respectively. we have shown that entanglement of the three-qubit
pure state can be locally or non-locally copied with the help of
local quantum copiers or non-local quantum copiers, respectively,
and that the amount of entanglements of the resultant copiers is
reduced. It has been found that it is more efficient to preserve
the entanglement by using non-local methods. The amount of
entanglement decreases rapidly with the increase of the number of
copying times in the iterating quantum copying. The amount of both
$E_3$ and $E_2$ entanglements approach  zero after six-times
iterating quantum copying. It means that even qubits, which are
copies of copies of copies of copies of copies of copies of the
GHZ state are already in local state. They do not have any
nonclassical correlations and are useless as a resource in
cryptographic conferencing or in multi-partite generalizations of
super-dense coding \cite{15}. This reduction is  due to a residual
entanglement between the copies and the quantum copying machine.
We have also shown that the inter-two-qubit entanglement is
preserved better than the inter-three-qubit entanglement in the
local cloning process while non-local cloning is much more
efficient than the local cloning for broadcasting entanglement,
and output state via non-local cloning exhibits the fidelity
better than local cloning.

\begin{center}
 {ACKNOWLEDGMENTS}
 \end{center}
 This work was supported in part by
the National Natural Science Foundation,  EYTF of the Educational
Department of China, and Hunan Province STF.


\begin{references}
\bibitem{1}  W. K. Wootters and W. H. Zurek, Nature 299 (1982) 802.
\bibitem{2}  D. Mozyrsky, V. Privman, and M. Hillery, Phys. Lett. A 226 (1997) 253.
\bibitem{3}  V. Buzek and M. Hillery, Phys. Rev. A 54 (1996) 1844.
\bibitem{4}  N. Gisin and S. Massar, Phys. Rev. Lett. 79 (1997) 2153.
\bibitem{5}  D. Bruss, A. Ekert, and C.Macchiavello, Phys. Rev. Lett. 81 (1998)2598.
\bibitem{6}  A. Einstein, B. Podolsky and N. Rosen, Phys. Rev. 47 (1935) 777;
             E. Schr\"{o}dinger, Naturwissenschaften, 23 (1935) 807; 823; 844.
\bibitem{7}  C. H. Bennett, G. Brassard, C. Cr\'{e}peau, R. Jozsa, A. Peres, and W. K. Wootters,
             Phys. Rev. Lett. 70 (1993) 1895;
             D. Bouwmeester, J.-W. Pan, K. Mattle, M.Eibl, H. Weinfurter, A.
             Zeilinger, Nature 390 (1997) 575.
\bibitem{8}  C. H. Bennett and S. J. Wiesner, Phys. Rev. Lett. 69 (1992) 2881.
\bibitem{9}  P. W. Shor, Phys. Rev. A 52 (1995) 2493.
\bibitem{10}  D. Deutsch, Proc. R. Soc. London, A 425 (1989) 73.
\bibitem{10'} T. Jennewein, C. Simon, G. Weihs, H. Weinfurter, and A. Zeilinger, Phys. Rev. Lett. 84 (2000) 4729.
\bibitem{10''} V. Giovannetti, S. LIoyd, and L. Maccone, Nature 412 (2001)417;
               V. Giovannetti,  S. Lloyd, and L. Maccone, Phys. Rev. A 65 (2002) 022309.
\bibitem{11}  P. Masiak and P. L. Klight, quant-ph/9808043.
\bibitem{12}  V. Buzek, V. Vedral, M. B. Plenio, P. L. Knight, and M. Hillery, Phys. Rev. A 55 (1997) 3327.
\bibitem{13'} J. Schlienz and G. Mahler, Phys. Rev. A 52 (1995) 4396.
\bibitem{13}  Z.-Y. Tong and L.-M. Kuang, Chin. Phys. Lett. 17 (2000) 469.
\bibitem{17}  V. Buzek and M. Hillery,   Phys.Rev.    A 54  (1996) 1844;
              D. Bru\ss, D.P. DiVincenzo, A. Ekert, C.A. Fuchs, C. Macchiavello, and J.A.
              Smolin, Phys. Rev.     A 57  (1998) 2368.
\bibitem{14}  V. Buzek and M. Hillery, quant-ph/9801009.
\bibitem{18}  R. F.  Werner,  Phys. Rev. A58 (1998) 1827.
\bibitem{15}  S. Bose, V. Vedral and P. L. Klight, Phys. Rev. A. 57 (1998) 822.
\end{references}
\end{document}